\documentclass[12pt]{iopart}

\usepackage{natbib}
\usepackage{hyperref}
\usepackage{url}
\usepackage{graphicx}
\usepackage{subcaption}
\usepackage{amssymb}
\usepackage{bbm}
\usepackage{upgreek}
\usepackage{xcolor}

\makeatletter
\def\@evenhead{\hfil \thepage} 
\def\@oddhead{\hfil \thepage}  
\makeatother

\begin{document}

\title{Deep Probabilistic Direction Prediction in 3D with Applications to Directional Dark Matter Detectors}

\author{Majd Ghrear, Peter Sadowski, Sven E. Vahsen \\
University of Hawai`i at M\=anoa\\
\texttt{\{majd,peter.sadowski,sevahsen\}@hawaii.edu} \\
}

\begin{abstract}
We present the first method to probabilistically predict 3D direction in a deep neural network model. The probabilistic predictions are modeled as a heteroscedastic von Mises-Fisher distribution on the sphere $\mathbb{S}^2$, giving a simple way to quantify aleatoric uncertainty. This approach generalizes the cosine distance loss which is a special case of our loss function when the uncertainty is assumed to be uniform across samples. We develop approximations required to make the likelihood function and gradient calculations stable. The method is applied to the task of predicting the 3D directions of electrons, the most complex signal in a class of experimental particle physics detectors designed to demonstrate the particle nature of dark matter and study solar neutrinos. Using simulated Monte Carlo data, the initial direction of recoiling electrons is inferred from their tortuous trajectories, as captured by the 3D detectors. For $40\,$keV electrons in a $70\%$ $\textrm{He}$ $30 \%$ $\textrm{CO}_2$ gas mixture at STP, the new approach achieves a mean cosine distance of $0.104$ ($26^\circ$) compared to $0.556$ ($64^\circ$) achieved by a non-machine learning algorithm. We show that the model is well-calibrated and accuracy can be increased further by removing samples with high predicted uncertainty. This advancement in probabilistic 3D directional learning could increase the sensitivity of directional dark matter detectors.
\end{abstract}

%
%
%
%
%

\section{Introduction}

Machine learning models are often used to predict directions, such as the direction of a moving particle or the direction of flow in a vector field. Since directions reside in non-euclidean geometries, the problem becomes a special case of regression. Specifically, a direction in $n$-dimensions can be uniquely represented by a $n$-dimensional vector residing on the hypersphere manifold $\mathbb{S}^{n-1}=\{\mathbf{x}| \mathbf{x}^T\mathbf{x}=1\}$. Thus, 2D directions reside on the manifold $\mathbb{S}^1$ (a circle) and 3D directions reside on the manifold $\mathbb{S}^2$ (a sphere). Models that fail to account for these topologies discard an important source of implicit bias that helps generalization: the smoothness of the function they are trying to learn. In probabilistic models, accounting for topology is important for defining valid probability distributions. In this work, we present the first deep learning model for probabilistically predicting 3D direction. Our model accounts for the $\mathbb{S}^2$ geometry of the problem by predicting distributions that are parameterized by the von Mises-Fisher (vMF).

The vMF is among the most important distributions in the well-established field of directional statistics~\citep{mardia1972statistics,mardia2000directional,ley2018applied}. Previous work has used linear models to parameterize directional distributions (including the vMF)~\citep{Chang,Rivest,Rosenthal,paine2020spherical}. However, these \textit{shallow} models have limited capacity to fit complex data such as images or point clouds, and in this work we extend the approach to deep neural networks. As neural network training is typically performed using stochastic gradient descent on the log data likelihood, additional care must be taken to ensure the objective function is smooth and numerically stable.

In order to clarify our contribution, we summarize related work in three related areas. In Section~\ref{2ddir}, we review the literature on deep models for predicting direction using non-probabilistic methods or methods that do not use 3D geometry. In Section~\ref{3Dorient}, we review the literature for the related problem of probabilistically predicting 3D orientations. In Section~\ref{embedding}, we discuss the use of directional statistics for learning deep embedding spaces.

\begin{figure}[ht]
\begin{center}
\includegraphics[width=0.85\textwidth,trim={1cm 5cm 0.5cm 12cm},clip]{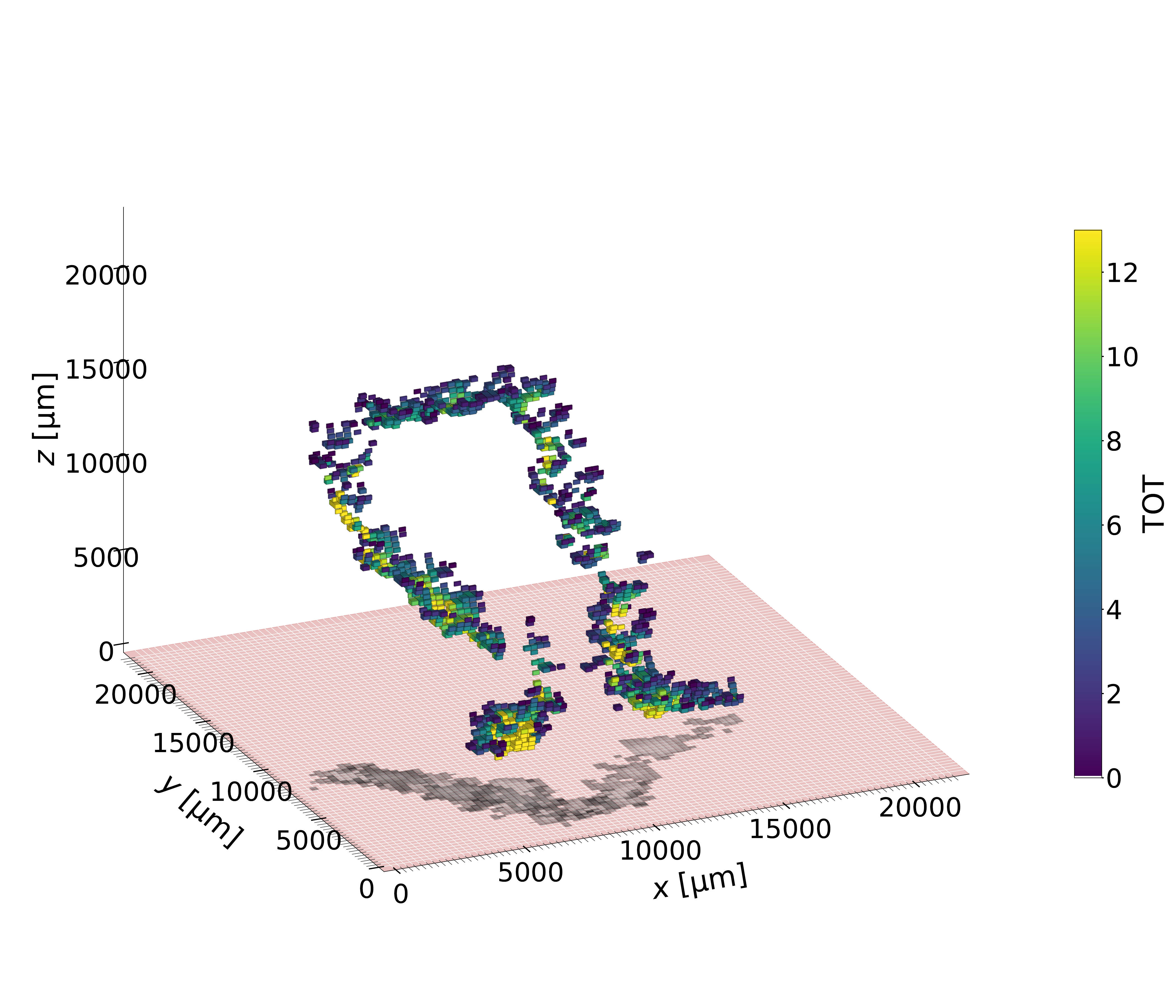}
\end{center}
\caption{An example of an experimentally detected $40\,$keV electron recoil, using the detector described in~\cite{Jaegle:2019jpx}. Electron recoils can have complex shapes and ambiguous starting points. An agent must identify the start of the recoil and determine how much of it to fit to determine the initial direction accurately. The color scale depicts a measure of ionization referred to as the time over threshold (TOT).}
\label{real_recoil}
\end{figure}

We demonstrate our approach on an application to directional dark matter detectors. These experiments detect elusive, electrically neutral elementary particles by observing the recoil trajectories of atomic nuclei or electrons produced when the neutral particles scatter in a detector. Figure~\ref{real_recoil} shows an example electron recoil detected in 3D, using the detector developed by~\cite{Jaegle:2019jpx}. This detector provides a highly-segmented 3D voxel grid reconstruction of the electron recoil trajectory, where each voxel is given a value corresponding to the amount of ionization detected at its position. Inferring the initial direction of the recoiling electron from this data is a difficult challenge. There are three main obstacles: First, the electron track is not straight, as the electron scatters in the gas. Second, the recoil starts and ends inside the detector, and it is not clear which side is the beginning, and which side is the end. Third, the non-straight trajectory and the start-versus-end ambiguity both make it difficult to estimate the uncertainty of any estimated direction. Unlike many common computer vision problems, even a trained human expert has difficulty correctly identifying the direction.

Through experiments we show that our probabilistic direction predictor significantly outperforms traditional algorithms when applied to a simulated electron recoil data set. Leveraging our framework’s uncertainty predictions, performance can be enhanced by discarding high-uncertainty samples. By discarding only the top 1\% of events with highest directional uncertainty, our model achieves the best expected directional performance (as defined in~\ref{NML_algo}). This development is far-reaching as it has the potential to enhance the sensitivity of all directional recoil detection experiments. Exciting possibilities include aiding in the discovery of dark matter~\citep{Billard:2013qya, OHare:2021utq} or resolving the solar abundance problem~\citep{ villante2019updated,OHare:2022jnx}. The framework can also be applied to general problems where direction needs to be predicted. We demonstrate this by applying it to the task of detecting the directions 3D arrows in~\ref{arrows}. 

Our contributions include: (1) The first deep learning framework that predicts 3D direction probabilistically, (2) a derivation of the negative log likelihood (NLL) loss function and approximations that stabilize training, (3) two experiments that empirically demonstrate learning and calibration, something which is lacking in related work, and (4) a deep model for predicting the direction of electron recoils from 3D point cloud data using sparse 3D convolutional networks.

\section{Related work}
\label{discuss}

\subsection{Deep models for predicting direction}
\label{2ddir}

While deep learning has been applied to direction prediction, the published methods are either not in 3D or not probabilistic.~\citet{GLASER2023102781} use a model with three outputs --- one for each Cartesian coordinate in 3D --- to predict the direction of neutrinos. An L2-normalization layer ensures the predicted direction lies on the unit sphere $\mathbb{S}^2$, but the model is not probabilistic. Previous work on determining the initial direction of electron recoils has been carried out in the context of X-ray polarimetry~\citep{WEISSKOPF20161179,PEIRSON2021164740}. However, this work is limited to 2D directions.~\cite{KITAGUCHI2019162389} probabilistically predicts 2D direction by splitting the unit circle into 36 bins and predicting the probability of each bin. There are two disadvantages to this method: (1) discretization reduces resolution, and (2) classification assumes the bins are unrelated and thus inductive bias is lost, a problem that becomes extreme as the number of bins increases. Thus, predictions in the form of a continuous probability density function are needed.~\citet{PEIRSON2021164740}, predicts 2D direction continuously using a Gaussian distribution. However, the von Mises distribution is more appropriate for this task as it accounts for the data residing on $\mathbb{S}^1$.

\cite{deepdirectstat2018} is the first deep probabilistic model to account for the fact that 2D directions reside on $\mathbb{S}^1$. Probabilistic predictions are expressed as von Mises distributions over a single angle.~\cite{deepdirectstat2018} also extends the framework to 3D orientations by jointly predicting distributions over three Euler angles (pan, tilt, and roll). However, this approach has a well-known drawback: Euler angles are susceptible to gimbal lock~\citep{hoag1963apollo}, a limitation acknowledged in subsequent studies, discussed in Section~\ref{3Dorient}. The framework of~\cite{deepdirectstat2018} could, in principle, be applied to probabilistic predictions of 3D direction by jointly predicting distributions over two angles ($\theta$ and $\phi$, spherical coordinates). Doing so would require special care in re-defining the von Mises distribution over a half circle, since $0 \leq \theta \leq \pi$. However, this approach would also suffer from gimbal lock (or coordinate singularities) as the $\phi$ degree of freedom becomes degenerate when $\theta = 0$. Our framework avoids this issue by predicting distributions on the manifold $\mathbb{S}^2$.  
  
\subsection{Deep models for predicting orientation}
\label{3Dorient}
Substantial work has been done in field of computer vision to develop methods for predicting the \textit{orientation} of a rigid body in 3D~\citep{LiaoCVPR19, huang2018cooperative, xiang2018posecnn, 8014807, Mahendran2018AMC}. Orientation is defined as the imaginary rotation needed to move an object from a reference placement to its current placement, disregarding translations. 3D orientation resides on SO(3) and is a different object from 3D direction which resides on $\mathbb{S}^2$. While~\cite{deepdirectstat2018} was the first to consider modeling orientation uncertainty, more recent approaches utilize representations other than Euler angles to circumvent gimbal lock.~\cite{Gilitschenski2020Deep} and~\cite{deng2022deep} represent 3D orientation with unit quaternions, predicting distributions over $\mathbb{S}^3$ parameterized by the antipodally symmetric Bingham distribution~\citep{Bingham1974AnAS}. Alternatively,~\cite{NEURIPS2020_33cc2b87} represents orientation with $3 \times 3$ rotation matrices, predicting distributions over SO(3) parameterized by the Matrix Fisher distribution~\citep{Khatri1977TheVM}. The approaches of~\cite{pmlr-v139-murphy21a},~\cite{liu2023delving}, and~\cite{klee2023image} address symmetric objects by predicting arbitrary distributions on SO(3).~\cite{pmlr-v139-murphy21a} represents distributions implicitly and then normalizes them whereas~\cite{liu2023delving} uses normalizing flows to directly generate normalized distributions. \cite{klee2023image} uses a Fourier basis of SO(3) to represent distributions and SO(3)-equivariant layers for sample-efficiency.

These probabilistic 3D orientation models yield distributions of rotations on SO(3), not directions on $\mathbb{S}^2$. Motivated by the definition of orientation, one could define a mapping $f:\textrm{SO}(3) \rightarrow \mathbb{S}^2$ as $f(\mathbf{R})= \mathbf{R} \cdot \mathbf{v}_R$. Here, $\mathbf{v}_R \in \mathbb{S}^2$ is a reference direction and $\mathbf{R} \in \textrm{SO(3)}$ is a rotation taking the reference direction to the truth direction. This provides a way to re-label truth directions as rotations ($\mathbf{R}$) with respect to a reference direction ($\mathbf{v}_R$). However, $f$ is not injective. SO(3) is a higher dimensional manifold than $\mathbb{S}^2$ and for any $\mathbf{v} \in \mathbb{S}^2$, there exists an uncountably infinite set $\{ \mathbf{R} \in \textrm{SO(3)} |  f(\mathbf{R}) = \mathbf{v} \}$. Probabilistic 3D orientation frameworks can not leverage the reduced dimensionality of 3D direction finding problems and must therefore learn these symmetries. This comes with practical implications. In training an orientation model for direction finding, the degeneracy of rotations corresponding to the truth direction means that an arbitrary choice must be made. Furthermore, those interested in probabilistic direction predictions need uncertainties expressed on $\mathbb{S}^2$, not SO(3). The fact that $f$ is not injective makes it difficult to extract a relevant distribution on $\mathbb{S}^2$. The probability density over any  $\mathbf{v} \in \mathbb{S}^2$ must be obtained as a non-trivial integral of the distribution on SO(3) over $\{ \mathbf{R} \in \textrm{SO(3)} |  f(\mathbf{R}) = \mathbf{v} \}$.

\subsection{Deep embedding models}
\label{embedding}

Hyperspheres have long been popular manifolds for embedding data. Embedding data on the curved hypersphere provides a natural way to balance the competing attraction of similar data and repellence of dissimilar data in clustering~\citep{banerjee2005clustering} or multidimensional scaling~\citep{elad2005texture}. For the same reason, many deep learning models constrain latent space representations to a hypersphere, for example in variational autoencoders~\citep{davidson2018hyperspherical} or in contrastive learning~\citep{he2020momentum,chen2020simple}. Some of this work makes use of the vMF distribution to quantify the variance in heteroscedastic predictions on the hypersphere latent space~\citep{davidson2018hyperspherical}. However, these embedding methods are \textit{unsupervised} in that the embeddings on the hypersphere are learned to optimize some objective. This work concerns the supervised learning scenario in which fixed targets live on the sphere.


\section{Probabilistic 3D Direction modelling with vMF}
\label{sec:proposed_model}

\subsection{The Loss Functions}

The cosine distance loss is a common non-probabilistic loss function for predicting directions. It is defined as
\begin{equation}
\label{CSloss}
        L_{cos} \triangleq \frac{1}{N}\sum_{i = 1}^N 1 - \mathbf{y}_i \cdot \hat{\mathbf{y}}_i,
\end{equation}
where $\mathbf{y}_i$ and $\hat{\mathbf{y}}_i$ are 3D unit vectors representing the true direction and predicted direction, respectively. The sum is over $N$, the number of examples in the training set. This can be generalized to a probabilistic model by defining a distribution over directions on $\mathbb{S}^2$. The 5-parameter Fisher–Bingham distribution~\citep{kent} is the analogue of the bivariate normal distribution on $\mathbb{S}^2$ with probability density function (p.d.f.)
\begin{equation}
\label{okent}
    f(\mathbf{x}) = \frac{1}{c(\kappa,\beta)} \exp{ \{ \kappa \mathbf{v}_1 \cdot \mathbf{x} + \beta \left[ (\mathbf{v}_2 \cdot \mathbf{x} )^2 - (\mathbf{v}_3 \cdot \mathbf{x})^2 \right] \} }.
\end{equation}
Above, $c(\kappa,\beta)$ is a normalization constant, $\kappa \in \mathbb{R}_+$ determines the spread, $\beta \in [0,\kappa / 2)$ determines the ellipticity, and $\mathbf{v}_1, \mathbf{v}_2, \mathbf{v}_3$ determine the mean direction, major axis, and minor axis, respectively. Equation~\ref{okent} can be simplified by assuming $\beta=0$, making the distribution isotropic. This yields the p.d.f.
\begin{equation}
\label{vMF}
    f(\mathbf{x}) = \frac{\kappa}{4 \pi \sinh{\kappa}} \exp{ \left( \kappa \mathbf{v}_1 \cdot \mathbf{x} \right) }.
\end{equation}
This special case of the 5-parameter Fisher–Bingham distribution is known as the von Mises–Fisher distribution on $\mathbb{S}^2$~\citep{mardia2000directional}.

Given the predicted directions, $\hat{\mathbf{y}}_i$, the predicted uncertainty parameters, $\kappa_i$, and the true directions $\mathbf{y}_i$, we use Equation~\ref{vMF} to calculate the likelihood as
\begin{equation*}
    \mathcal{L} = \prod_{i=1}^N \frac{\kappa_i}{4 \pi \sinh{\kappa_i}} e^{\kappa_i (\mathbf{y}_i\cdot \hat{\mathbf{y}}_i )},
\end{equation*}
and the NLL, scaled by batch size, is
\begin{equation}
\label{NLL}
    NLL = \frac{-1}{N}\sum_{i = 1}^N  \ln(\frac{\kappa_i}{4 \pi \sinh{\kappa_i}}) + \kappa_i ( \mathbf{y}_i \cdot \hat{\mathbf{y}}_i ).
\end{equation}
We use this as the loss function for our probabilistic direction prediction model. If we make an additional assumption that $\kappa_i$ is constant over $i$ (the homoscedastic assumption), the first term in Equation~\ref{NLL} becomes a constant and the optimization problem becomes equivalent to minimizing the cosine distance (Equation~\ref{CSloss}). 

\subsection{Practical considerations}
\label{Pcon}
Some considerations are required to make the likelihood function and gradients of the probabilistic model tractable. First, the $g(\kappa_i) \equiv \ln{\frac{\kappa_i}{4 \pi \sinh{(\kappa_i)}}}$ term on the right-hand side of Equation~\ref{NLL} is numerically unstable. When $\kappa_i = 0$, the vMF distribution in Equation~\ref{vMF} is uniform over the sphere, but during learning $\kappa_i$ generally increases as the distribution becomes more localized (uncertainty decreases). If $\kappa_i \geq 87$, the denominator of $g(\kappa_i)$ evaluates to zero (assuming a 32-bit floating point data type) resulting in an infinite loss. To avoid this, we use the limit

\begin{centering}
\begin{equation*}
    \lim_{ \kappa_i \to\infty} g(\kappa_i) = \ln(\frac{\kappa_i}{2\pi}) - \kappa_i.
\end{equation*}
\end{centering}

In fact, $g(\kappa_i) \approx \ln(\frac{\kappa_i}{2\pi}) - \kappa_i$ is a well-known approximation even for small values of $\kappa_i$~\citep{dhillon2003modeling}. For example, if
$\kappa = 9$, the values of $\ln(\frac{\kappa_i}{4 \pi \sinh{(\kappa_i)}})$ and $\ln(\frac{\kappa_i}{2\pi}) - \kappa_i$ are indistinguishable when using the 32-bit floating point data type. Therefore, we replace the loss definition from Equation~\ref{NLL} with
\begin{equation}
\label{try1}
    \textrm{NLL} = \frac{-1}{N}\sum_{i = 1}^N  \mathbbm{1}_{\kappa_i<9}  \left( g(\kappa_i) \right) +  \mathbbm{1}_{\kappa_ \geq 9} \left( \ln\left(\frac{\kappa_i}{2\pi}\right) - \kappa_i \right)  + \kappa_i ( \mathbf{y}_i \cdot \hat{\mathbf{y}}_i ).
\end{equation}

This function can be implemented in \texttt{PyTorch}~\citep{paszke2019pytorch} using the \texttt{torch.where} function, but unfortunately at the time of this writing,  \texttt{PyTorch} produces a \texttt{nan} gradient if either of the terms evaluates to infinity~\citep{paszke2019pytorch}. Thus we solve this problem by removing the term $g(\kappa_i)$ from Equation~\ref{try1} completely, approximating it with its 15th-order Taylor Series about $\kappa_i = 0$
\begin{equation*}
    T_{15} (\kappa_i) \triangleq \sum^{15}_0 \frac{g^{(n)}(0)}{n!} \kappa_i^n =  \ln(4\pi) + \frac{\kappa^{2}}{6} - \frac{\kappa^4}{180} + \frac{\kappa^6}{2835} - \frac{\kappa^8}{37800} + \dots .
\end{equation*}
We stop at the 15th-order because higher powers of $\kappa$ make the Taylor series unstable. We now have two approximations for $\ln{\frac{\kappa_i}{4 \pi \sinh{(\kappa_i)}}}$, for low and high $\kappa$. Putting these together we arrive at the final form of the loss function for the model
\begin{equation}
    \label{Floss}
    \textrm{NLL} = \frac{-1}{N}\sum_{i = 1}^N \texttt{torch.where} \left( \kappa_i < 2.65, T_{15} (\kappa_i) ,  \ln(\frac{\kappa_i}{2\pi}) - \kappa_i \right)  + \kappa_i ( \mathbf{y}_i \cdot \hat{\mathbf{y}}_i ).
\end{equation}
The condition $\kappa_i < 2.65$ is chosen so that Equation~\ref{Floss} is continuous and is in agreement with Equation~\ref{NLL} within a margin of $0.14 \%$.

\section{Application to Electron Recoils}
\label{e_recoils}
For a demonstration of our method on a simple toy problem see~\ref{arrows}. Here, we apply it to directional dark matter detectors. The most mature directional recoil detection technology is the gas time projection chamber (TPC). Gas TPCs are capable of determining 3D direction of recoiling atomic nuclei or electrons by reconstructing their ionization tracks. The detectors produce a 3D voxel grid, where each voxel provides a measurement of ionization at that position, as illustrated in Figure~\ref{real_recoil}. We focus on determining the initial direction of the electron recoil signal because it represents the most complex scenario. There is also abundant literature on a non-machine learning algorithm for determining the direction of electrons in gas detectors~\citep{WEISSKOPF20161179,DiMarco_2022} which serves as a well-studied non-machine learning baseline in our experiments.

As an electron recoils in gas it undergoes multiple scattering which alters the trajectory of the recoiling electron and creates a track of secondary, ionized electrons. By reconstructing the positions of the secondary electrons in 3D, directional gas TPCs can infer the initial direction of the recoiling electron. Determining the direction is a complicated task because of the non-trivial track shapes, illustrated in Figure~\ref{real_recoil}. Typical (non-statistical) algorithms approach this by first predicting which side of the recoil track is the starting point, and then how much of the track beyond the starting point to associate with the initial trajectory. As the electron recoils it loses energy and in the process becomes more highly ionizing. Hence the charge density along the track can be used to determine the starting point. To first order, there are two effects influencing the angular resolution of electron recoils: the multiple scattering of the recoiling electron and the effective point resolution with which the secondary electrons are detected. Multiple scattering makes the electron lose directionality as it travels, suggesting only the very beginning of the track should be used to determine the initial direction. On the other hand, the effective point resolution sets a lower limit on the length scale where meaningful information can be extracted from the track. An agent needs to learn the best trade-off between these two effects on a track-by-track basis to provide the most accurate initial direction predictions.

\subsection{Simulation, Preprocessing, and Data splits}
\label{datasets}

\begin{figure}[ht]
    \begin{subfigure}{.49\textwidth} 
        \centering
        \includegraphics[width=\linewidth,trim={1cm 0.55cm 1cm 0.75cm},clip]{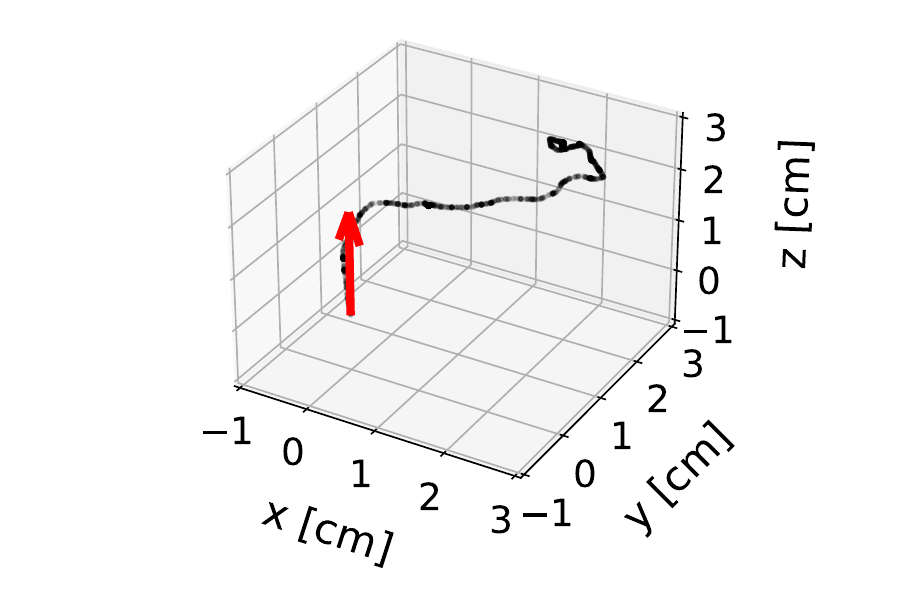}  
        \caption{Raw \textsc{Degrad} simulation}
        \label{sim1}
    \end{subfigure}
    \begin{subfigure}{.49\textwidth}
        \centering
        \includegraphics[width=\linewidth,trim={1cm 0.55cm 1cm 0.75cm},clip]{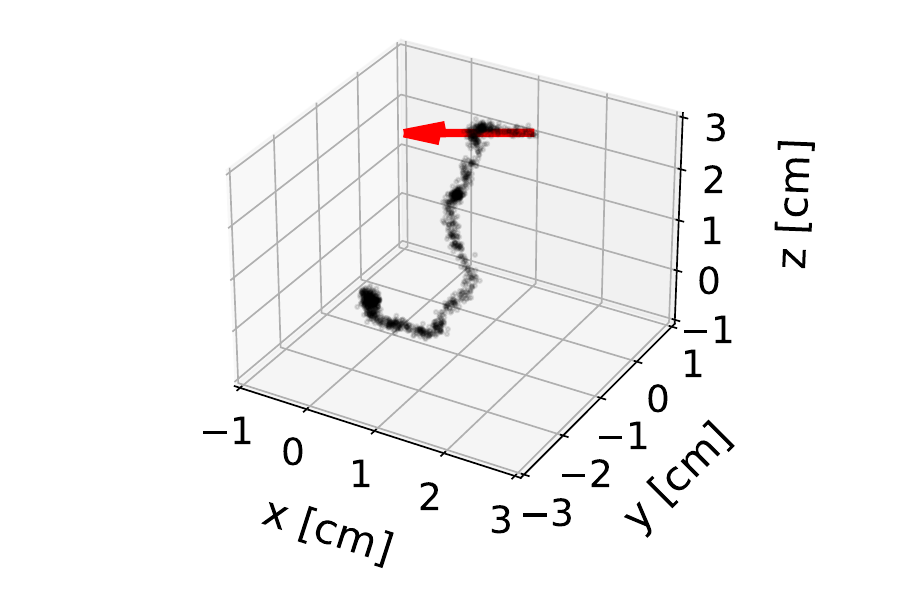}  
        \caption{Processed simulation}
        \label{sim2}
    \end{subfigure}\\
    \centering
    \begin{subfigure}{0.5\textwidth}
        \centering
        \includegraphics[width=\linewidth,trim={1cm 1cm 0cm 3.5cm},clip]{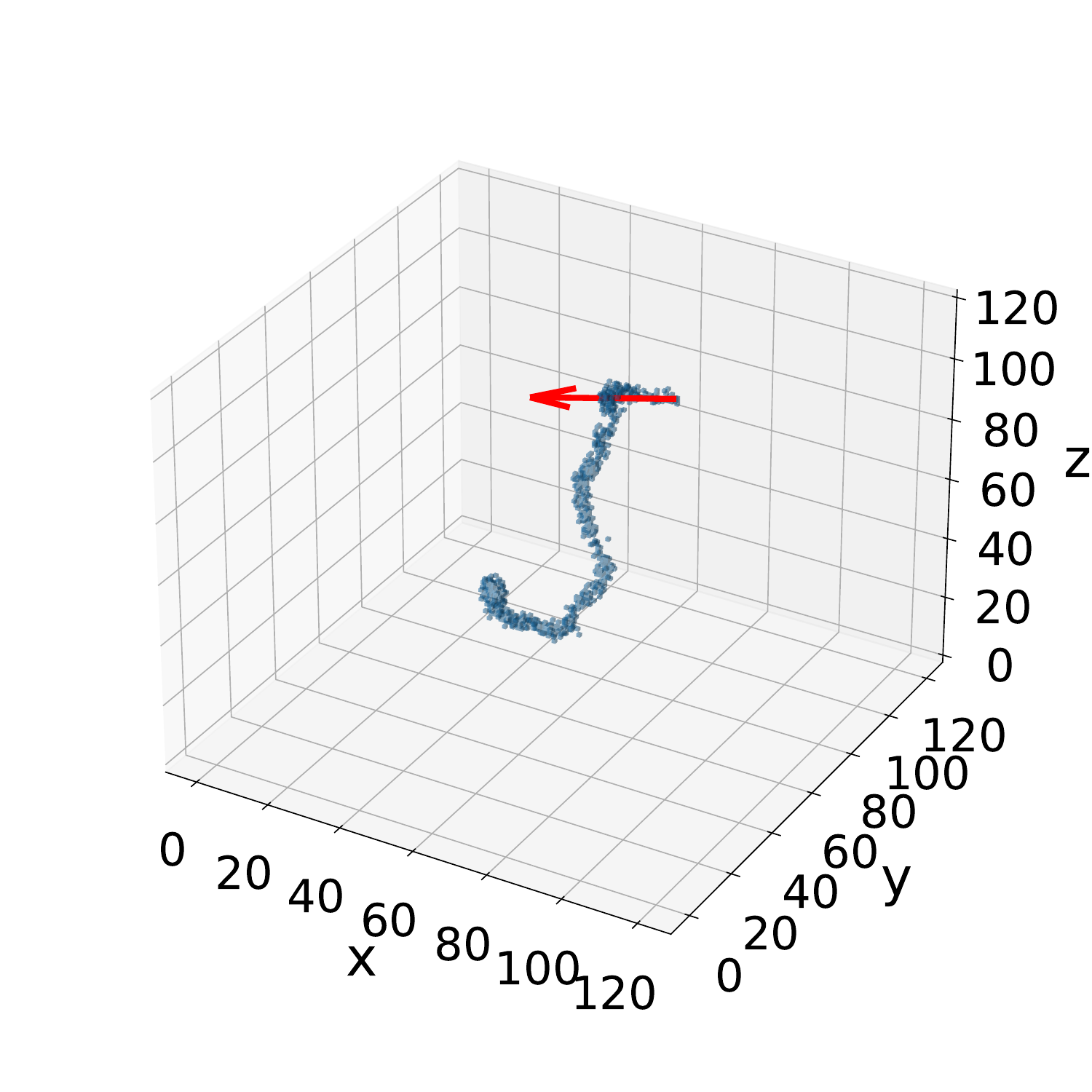} 
        \caption{Voxelized simulation}
        \label{sim3}
    \end{subfigure}
    \caption{An electron recoil simulation at various prepossessing stages. The raw \textsc{Degrad} simulation (a) is processed by randomly rotating, applying Gaussian smearing, and mean-centering (b). The point cloud is then binned into a $(120,120,120)$ voxel grid with a $(500\,{\upmu}{\rm m})^3$ resolution, simulating the finite resolution of the detector (c). The red arrows indicate the true direction.}
\end{figure}

Below, we detail the steps taken to create our simulated electron recoil data sets. The specification of the gas mixture, pressure, temperature, and diffusion are inspired by~\cite{Jaegle:2019jpx}.
\begin{enumerate}
    \item We simulate $10^6$ electron recoils at $40$, $45$, and $50\,$keV in a $70 \%$ He : $30\%$ $\textrm{CO}_2$ gas mixture at $20^\circ$C and $760$ Torr using \textsc{Degrad}~\citep{degrad}. All electron recoil simulations begin at the origin with the initial momentum in the positive z-direction. This produces a track of secondary (ionized) electrons for each electron recoil simulation, illustrated in Figure~\ref{sim1}.
    \item To model the diffusion in the detector, Gaussian smearing is applied to each ionized electron in the track. For each track, the amount of smearing is drawn from a uniform distribution of $160 - 466\,{\upmu}{\rm m}$ , the smallest and largest expected values in~\cite{Jaegle:2019jpx}.
    \item To make our simulations isotropic, a random rotation is applied to the track. The true initial direction after rotation is saved.
    \item The track is translated so that the origin is the center of charge. An example of a simulation at this stage is shown in Figure~\ref{sim2}
    \item The secondary electrons are then binned into a $(120,120,120)$ voxel grid, each voxel is a $(500\,{\upmu}{\rm m})^3$ cube. In binning, data is directly transformed into \texttt{PyTorch Sparse Tensors} in the COO format. Tracks that are not fully contained in the voxel grid are discarded. An illustration of a simulation at this stage is shown in Figure~\ref{sim3}.
\end{enumerate}
The final dataset contains 2,766,798 simulations. A random data split is used to create a training set (80\%) and a validation set (20\%). To create the test data sets we simulate another $2 \times 10^4$ electron recoils at $40$ and $50\,$keV. We follow the same steps outlined above except for a slight modification. The applied Gaussian smearing is no longer drawn from a uniform distribution. Instead, we probe two specific cases: a high diffusion case where $443\,{\upmu}{\rm m}$ of smearing is applied and a low diffusion case where $200\,{\upmu}{\rm m}$ is applied. The two energies and two diffusion cases give us four test data sets. Training our models across a range of energy and diffusion makes them more robust to variations in simulation parameters. Testing in specific scenarios, like high diffusion/low energy, evaluates performance in challenging conditions, while low diffusion/high energy evaluates performance in easier conditions. The specific test cases also simplify future comparisons.

The simulation framework, \textsc{Degrad}, is the standard software package used in physics for simulating the interaction of low-energy electrons in gasses~\citep{degradbest}. The focus of our work is to compare the different models, and our simulations provide a typical setup on which to compare performance. The performance in any particular detector is expected to differ slightly due to the sim-to-real gap. In the future we plan to use experimental radioactive source data to train our models, thereby avoiding the sim-to-real gap entirely.  Simulation parameters, such as diffusion and binning, were deliberately conservative compared to what is experimentally achievable.

\subsection{Architecture}
\label{arch}
The input to our 3D direction prediction model is a (120,120,120,1) voxel grid, which means every event has 1,728,000 features. The fraction of non-zero features (i.e. voxels with non-zero detected charge) for a typical electron recoil simulation is on the order of $10^{-4}$. Sparsity is a common feature in highly-segmented 3D data and it is essential to take advantage of it to keep computational requirements feasible. Data is stored as \texttt{PyTorch Sparse Tensors} in the COO(rdinate) format, where only the non-zero entries and their indices are stored. A DataLoader is used to load batches of sparse tensors on the fly. The coordinates and values of the non-zero features are passed into our models, where they are immediately converted into a \texttt{spconv.SparseConvTensor}~\citep{spconv2022}.

\begin{figure}[ht]
\begin{center}
\includegraphics[width=\textwidth,trim={0cm 0.4cm 0cm 0.55cm},clip]{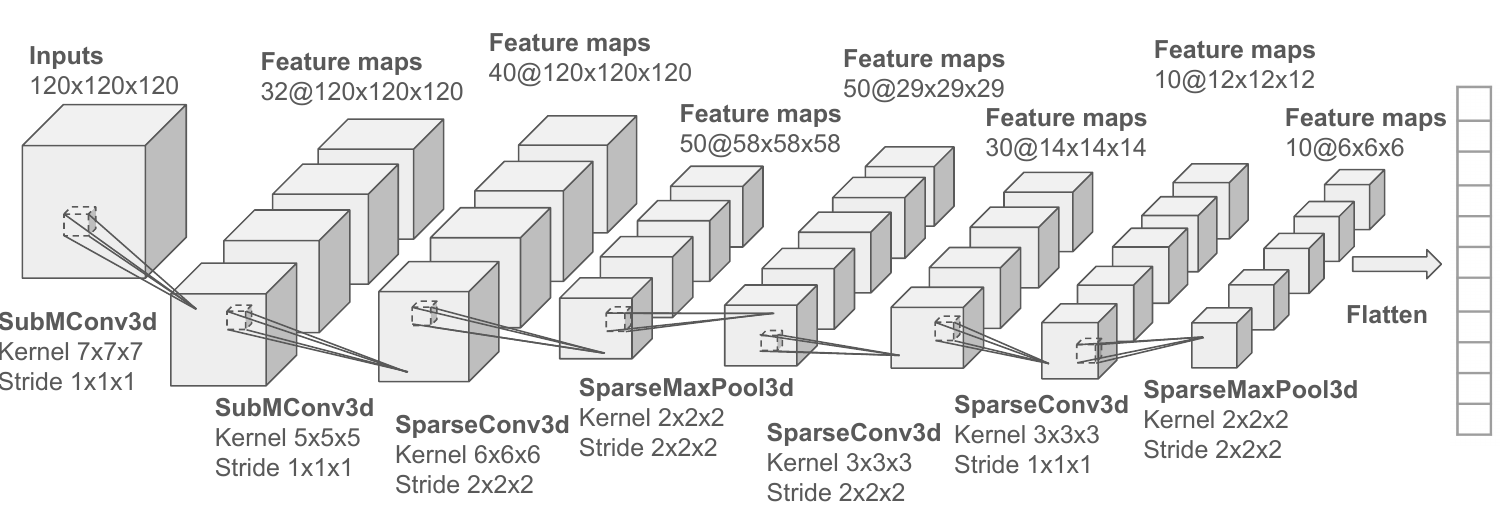}
\end{center}
\caption{The feature extraction portion of our models. Each convolutional layer includes a learnable bias and is followed by a RelU activation. Padding is only applied in the SubMConv3d layers, where it is set such that the input and output have the same shape. The implementation is via \texttt{SpConv}~\citep{spconv2022}.}
\label{Feature-ext}
\end{figure}

The model begins with a feature extraction portion, a series of submanifold sparse convolution, sparse convolution, and sparse max pooling layers. The details of the layers and the order in which they are applied is outlined in Figure~\ref{Feature-ext}. These sparse operations allow us to convolve and down-sample our input data without expressing them as dense tensors, significantly increasing speed and reducing memory usage. Sparse convolution and max pooling are equivalent to their dense counterparts, except they operate on sparse tensors. In submanifold sparse convolution~\citep{SubmanifoldSparseConvNet, 3DSemanticSegmentationWithSubmanifoldSparseConvNet}, padding is applied so that the input and output have the same shape. An output site is active if and only if the corresponding input site is active, in which case the output feature vector is computed in the same manner as for regular convolution. Since the number of active sites is unchanged, submanifold sparse convolution allows us to process our input through several layers while maintaining its sparseness. The implementation of the layers specified in Figure~\ref{Feature-ext} is through \texttt{SpConv}~\citep{spconv2022} which is based on~\cite{s18103337}.

\begin{figure}[ht]
\begin{center}
\begin{subfigure}{0.5\textwidth}
  \centering
  \includegraphics[width=\linewidth,trim={1cm 0.5cm 0.5cm 0.5cm},clip]{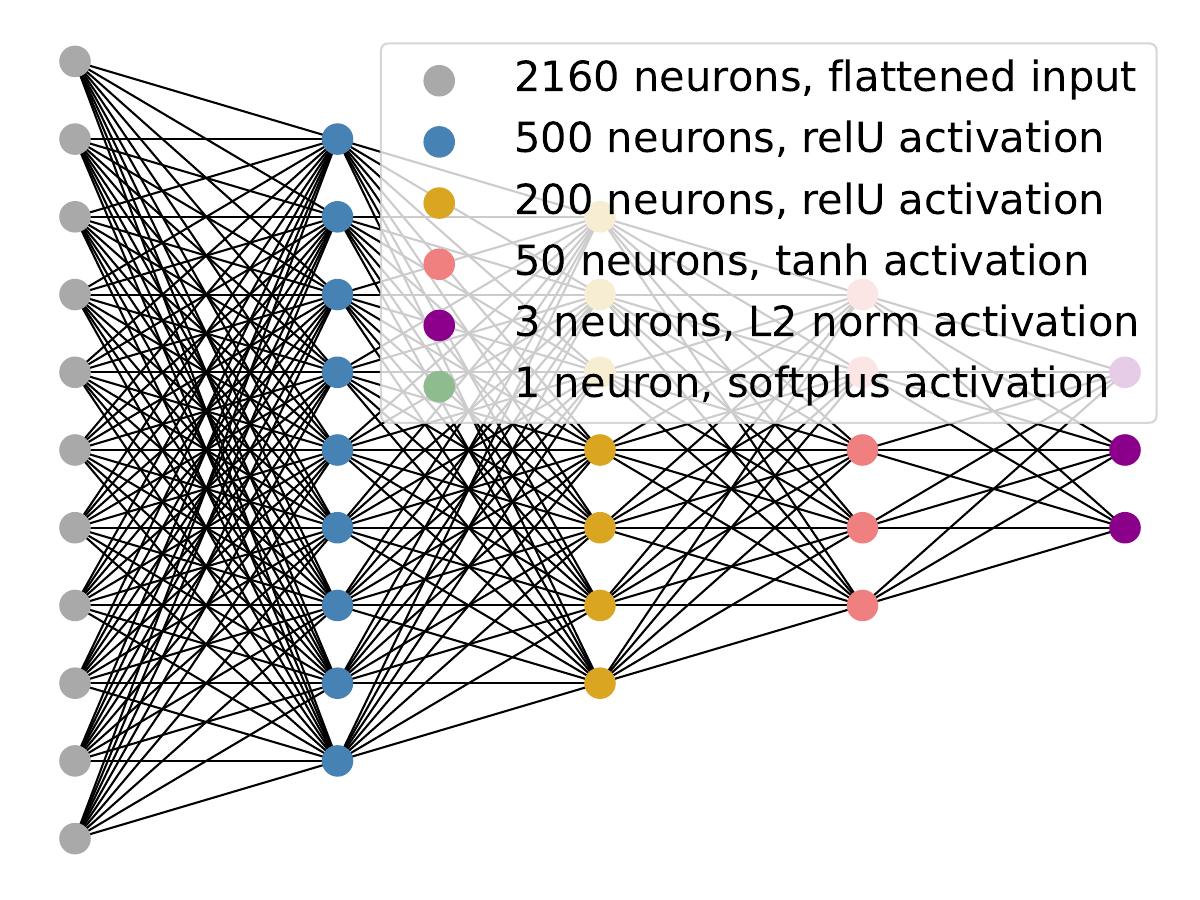}
\end{subfigure}%
\begin{subfigure}{0.5\textwidth}
  \centering
  \includegraphics[width=\linewidth,trim={1cm 0.5cm 0.5cm 0.5cm},clip]{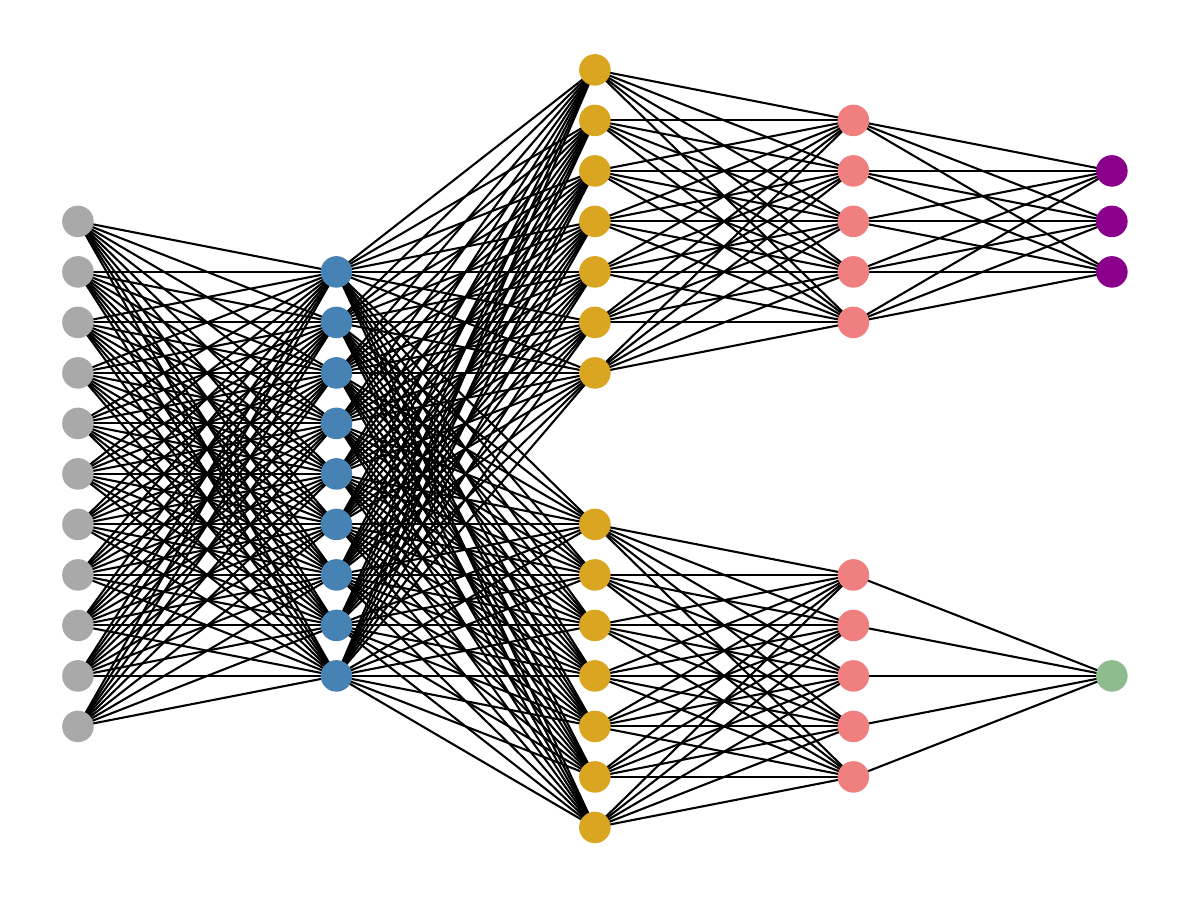}
\end{subfigure}
\end{center}
\caption{The dense portion of our models. The feature extraction described in Figure~\ref{Feature-ext} is followed by either the left or the right neural network to make the Det-NN or vMF-NN model, respectively. The illustrations do not display the actual number of neurons per layer. The number of neurons and the activation functions are specified in the legend. A learnable bias is added to every layer.}
\label{network_heads}
\label{NN_heads}
\end{figure}

The output of the 3D convolution layers is flattened and then passed to a sequence of densely-connected layers that produce the parameters of the vMF distribution (Figure~\ref{network_heads}b). One arm ends in three linear neurons with an L2-normalization activation function, interpreted as the predicted direction on $\mathbb{S}^2$ ($\hat{\mathbf{y}}_i$), while a second arm ends with a single output neuron with a Softplus activation function, interpreted as the predicted uncertainty ($\kappa_i$). To evaluate the advantage of the probabilistic approach, we also train a non-probabilistic version in which the uncertainty arm is removed (Figure~\ref{network_heads}a). We refer to the probabilistic, heteroscedastic convolutional network as vMF-NN and this non-probabilistic, ``deterministic'' convolutional network as Det-NN. The Det-NN baseline can be viewed as the 3D generalization of the biternion neural networks introduced by~\cite{beyer}. This represents the current state-of-the-art method for predicting direction in 3D.

\subsection{Training}

The Det-NN and vMF-NN models are trained by minimising the loss functions outlined in Equations~\ref{CSloss} and~\ref{Floss}, respectively, using the \texttt{Adam} optimizer~\citep{kingma2014adam}. The learning rate is set to ${\textrm{lr}=}0.0001$, the beta values are $(0.94, 0.999)$, and $\textrm{eps}= 10^{-7}$, while all other parameters are kept at the \texttt{PyTorch} default value. The deep learning models are trained with mini-batches of size $N=256$. Hyperparameters were optimized by exploring values in the range $\textrm{eps} \in [10^{-8}, 10^{-6}]$, $\textrm{lr} \in [10^{-6},10^{-4}]$, and $N \in [64,256]$. 

Both the Det-NN and vMF-NN models are trained on the training set while their loss on the validation set is used for early stopping and hyperparameter selection. If the validation loss has not decreased in the last five epochs, training is stopped. The weights corresponding to the lowest validation loss are saved and used for the final model.

\subsection{Performance}

We evaluate the trained deep learning models on the independent test sets discussed in Section~\ref{datasets} by comparing them to a commonly used non-ML algorithm and to our estimation of the best possible performance under certain idealized conditions. The non-machine learning algorithm is adapted from~\cite{DiMarco_2022} and generalized to 3D (Non-ML standard). This algorithm has three hyperparameters, so we also explore a variation (Non-ML tuned) in which the hyperparameters are fit to the test sets. The best expected directional performance is estimated by utilizing information that is only available in simulation, including the true starting point and the correct head-tail information (Best-Expected). The purpose of the Best-Expected metric is not to be a viable option for analysis but to provide a benchmark of exceptional performance. The Non-ML and Best-Expected algorithms are detailed in~\ref{NML_algo}.

\begin{figure}[ht]
\begin{center}
\begin{subfigure}{0.5\textwidth}
  \centering
  \includegraphics[width=\linewidth]{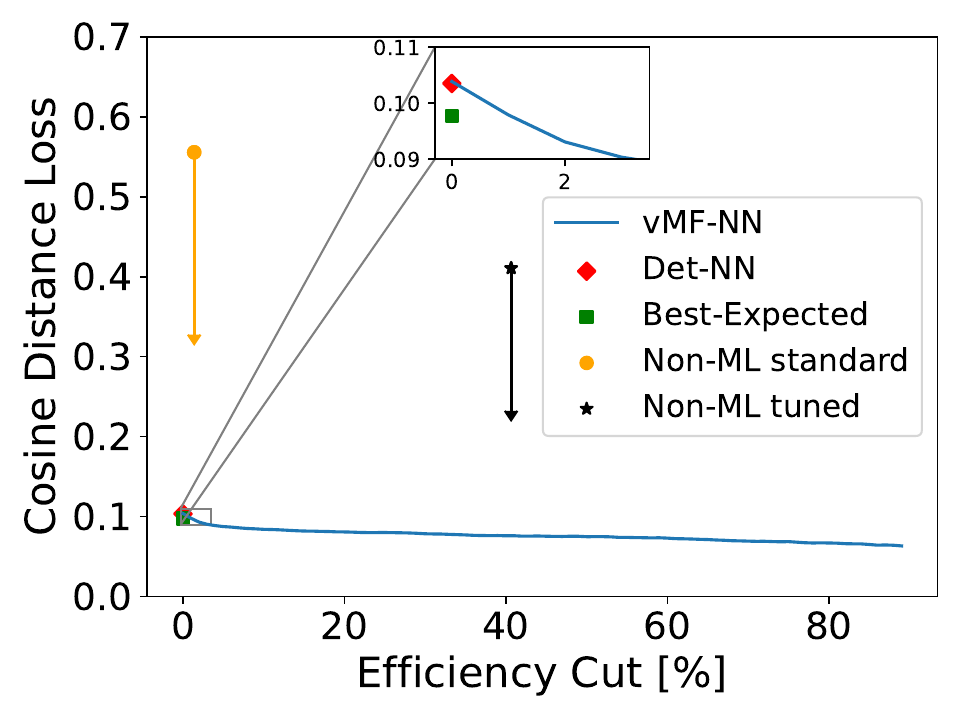}
\end{subfigure}%
\begin{subfigure}{0.5\textwidth}
  \centering
  \includegraphics[width=\linewidth]{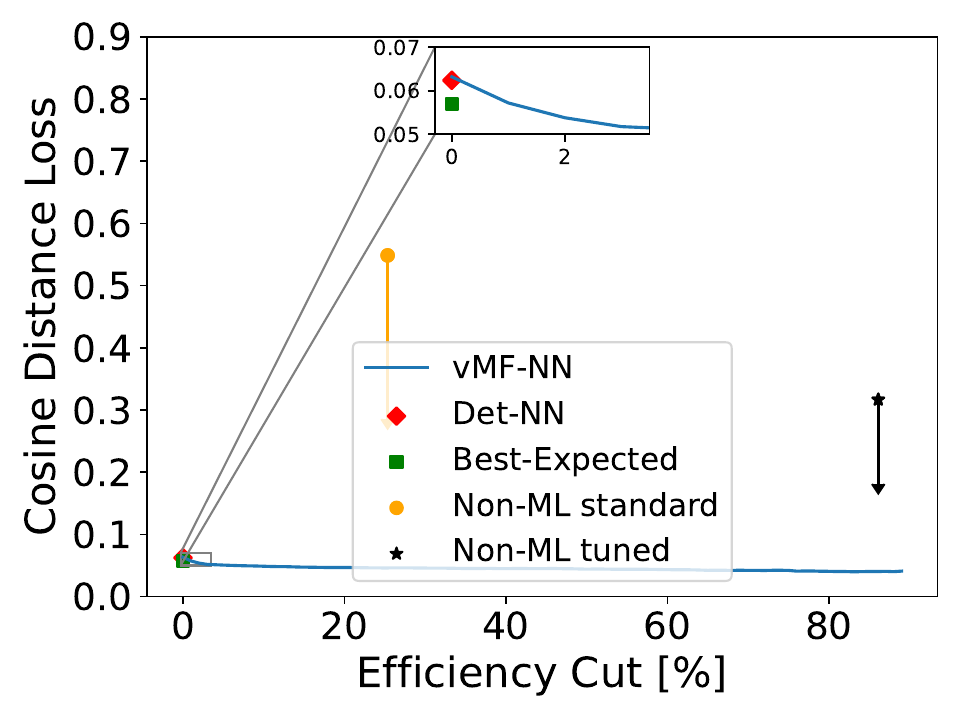}
\end{subfigure}
\end{center}
\caption{Cosine distance loss versus efficiency cuts for all models on the $40\,$keV test data set with $443\,{\upmu}{\rm m}$ Gaussian smearing (left) and the $50\,$keV test data set with $200\,{\upmu}{\rm m}$ Gaussian smearing (right). The Det-NN (red diamond) and Best-Expected (green square) models have no efficiency cuts. For the vMF-NN model, we improve accuracy by excluding examples with higher predicted uncertainty, results are presented with a blue curve. The orange point and black star indicate the performance of the Non-ML method with standard parameters and with parameters tuned on the test set, respectively. The downward arrows indicate how much improvement is possible if we help the Non-ML method by providing it with the true head-tail.}
\label{test_dir}
\end{figure}

The cosine distance loss (Equation~\ref{CSloss}) is used to compare performance. Hence, when comparing the models we plot the cosine distance loss versus the percentage of omitted tracks, referred to as the efficiency cut. In Figure~\ref{test_dir}, we display the performance of all models on the $40\,$keV test data set with $443\,{\upmu}{\rm m}$ Gaussian smearing (left) and the $50\,$keV test data set with $200\,{\upmu}{\rm m}$ Gaussian smearing (right). In both cases, we find that the deep learning models significantly outperform Non-ML, even under the most optimistic assumptions. On the $40\,$keV test data set with $443\,{\upmu}{\rm m}$ Gaussian smearing and at a $0\%$ efficiency cut, the Det-NN, vMF-NN, and Best-Expected models achieve a cosine distance loss of $0.104$, $0.104$, and $0.098$, respectively. Converting these values to degrees by taking the arccos of $1-L_{cos}$ gives $26.4^\circ$, $26.4^\circ$, and $25.6^\circ$. On the $50\,$keV test data set with $200\,{\upmu}{\rm m}$ Gaussian smearing and at a $0\%$ efficiency cut, the models achieve a cosine distance loss of $0.0624$, $0.0632$, and $0.0569$, in the same order. Performing the same conversion to degrees gives $20.3^\circ$, $20.5^\circ$, and $19.4^\circ$. The vMF-NN and Det-NN models have similar performance. However, the accuracy of the vMF-NN model can be improved by discarding events with low $\kappa_i$ (high predicted uncertainty). The vMF-NN model outperforms Best-Expected with just a $1\%$ efficiency cut, in both cases.

\begin{figure}[ht]
    \begin{subfigure}{.49\textwidth} 
        \centering
        \includegraphics[width=\linewidth]{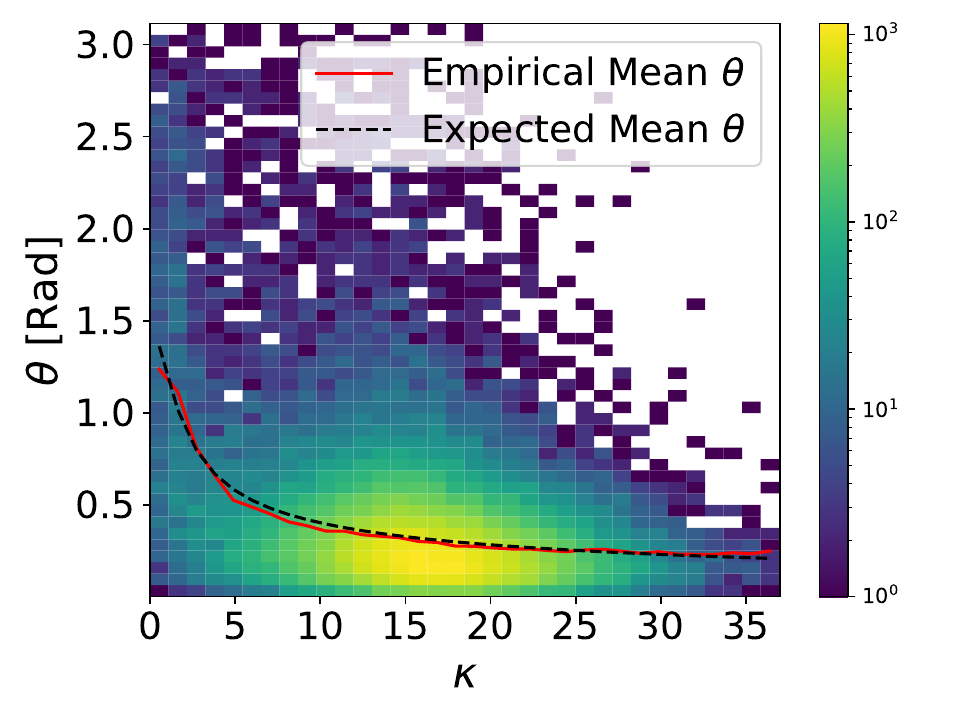}  
        \caption{ }
        \label{2dhist}
    \end{subfigure}
    \begin{subfigure}{.49\textwidth}
        \centering
        \includegraphics[width=\linewidth]{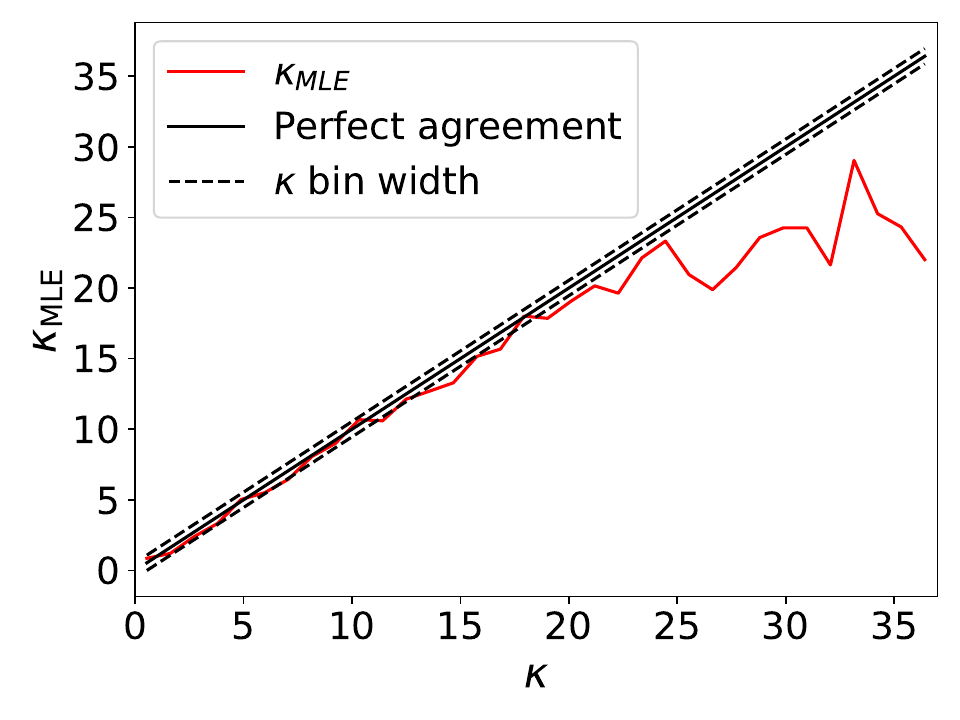}  
        \caption{ }
        \label{kappas}
    \end{subfigure}
    \caption{Left: 2D histogram of predicted uncertainty ($\kappa$) and the angle from true to predicted direction for all test data. The red curve is the mean angle for each bin in $\kappa$. The black dashed curve is the expected mean angle according to Equation~\ref{vMF}. Right: For each bin in $\kappa$ on the left hand side, the angular mismeasurements are used to find the MLE of $\kappa$ in Equation~\ref{vMF}. The MLE values are plotted versus the bin center $\kappa$ value.}
\end{figure}

To assess the calibration of the vMF-NN model we plot a 2D histogram of the predicted uncertainty ($\kappa_i$) and the angle from the true direction to the predicted direction ($\theta_i$) for all test data, displayed in Figure~\ref{2dhist}. The spread in $\theta_i$ gets smaller for higher values of $\kappa$, indicating that the vMF-NN model is predicting $\kappa_i$ appropriately. The solid red curve in Figure~\ref{2dhist} shows the mean angle, calculated for each bin in $\kappa$. Using Equation~\ref{vMF}, we can also calculate the expected mean angle as a function of $\kappa$
\begin{equation*}
    \theta_{\rm avg.} = \int_{0}^{2\pi} \int_{0}^{\pi} \theta \frac{\kappa \exp{ \left( \kappa \cos{(\theta)} \right) } }{4 \pi \sinh{\kappa}}   \sin(\theta) d\theta d\phi = \frac{\pi \left( I_o(\kappa) - e^{-\kappa} \right) }{2 \sinh(\kappa)},
\end{equation*}
where $I_o$ is a modified Bessel function of the first kind. The expected mean angle as a function of $\kappa$ is plotted as the dashed black line in Figure~\ref{2dhist}. The agreement of the expected mean angle with the mean angle for each bin in $\kappa$ suggests that the vMF-NN model is well-calibrated.

Building onto the 2D histogram in Figure~\ref{2dhist}, we check whether the predicted angular mismeasurements in a $\kappa$ bin behave as if they are drawn from a von Mises-Fisher distribution with that value of $\kappa$. For each bin in $\kappa$, we compute a distribution of $\cos{\theta_i}$ and fit it with Equation~\ref{vMF} to obtain the maximum likelihood estimation (MLE), $\kappa_{\rm{MLE}}$. In a perfect scenario, we expect $\kappa_{\rm{MLE}}$ to match the bin center value. The $\kappa_{\rm{MLE}}$ value for each bin versus the bin center value is plotted in Figure~\ref{kappas}. The diagonal solid line indicates perfect agreement and the dashed lines around it show the width of a bin in $\kappa$. Looking at both Figures~\ref{2dhist} and~\ref{kappas}, we see good agreement where the majority of our statistics lie. The discrepancy at higher values of predicted $\kappa$ indicates that the model is overconfident in its predictions. This is not surprising, given the known tendency for neural networks to overfit and be overconfident. However, this represents only a small fraction of the dataset.

\begin{figure}[ht]
    \begin{subfigure}{.49\textwidth} 
        \centering
        \includegraphics[width=\linewidth]{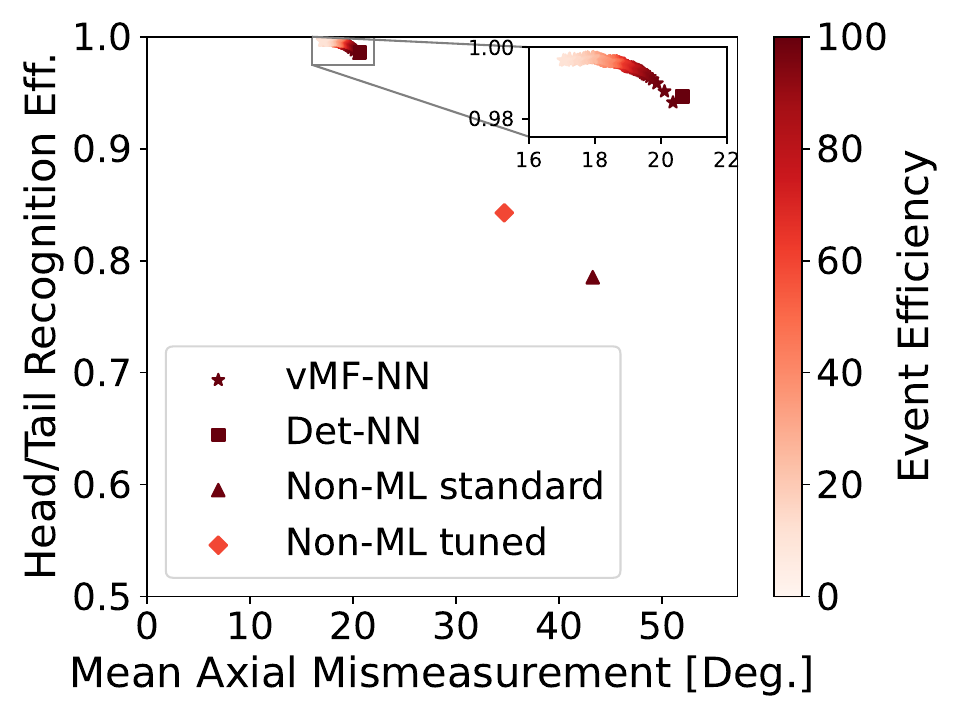}  
        \caption{ }
        \label{svensplot1}
    \end{subfigure}
    \begin{subfigure}{.49\textwidth}
        \centering
        \includegraphics[width=\linewidth]{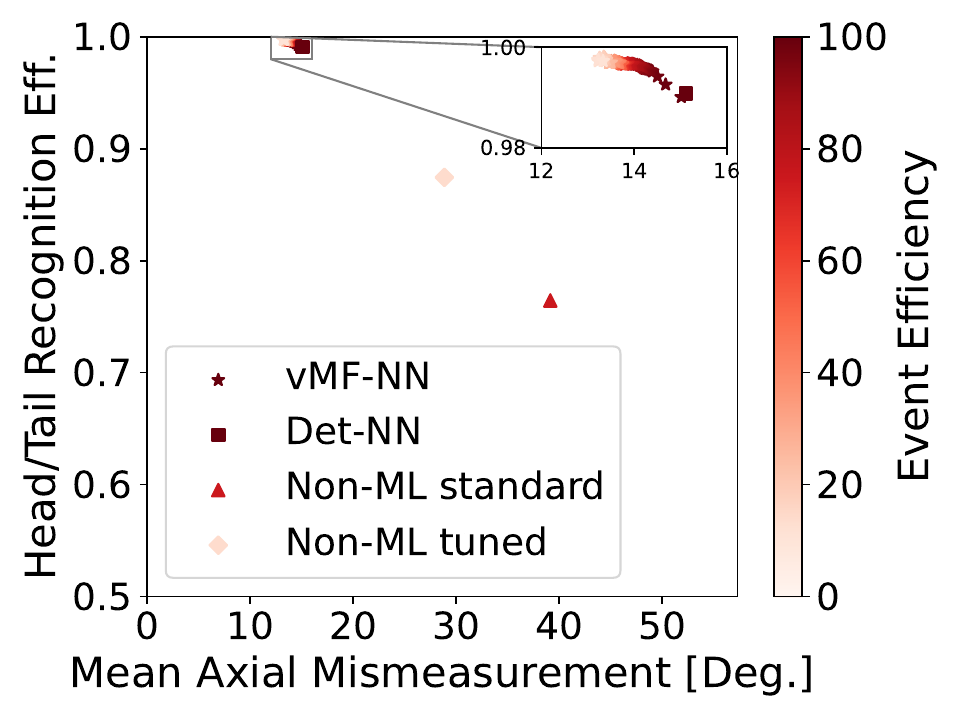}  
        \caption{ }
        \label{svensplot2}
    \end{subfigure}
    \caption{Left: Head/tail recognition efficiency versus mean axial mismeasurement for all models on the $40\,$keV test data set with $443\,{\upmu}{\rm m}$ Gaussian smearing. The color scale indicates the event efficiency. The vMF-NN and Det-NN models have similar performance at 100\% event efficiency. By omitting events with high predicted uncertainty, the performance of vMF-NN can be improved, as displayed by the curve on the top-left of the plot. Right: The same plot for the $50\,$keV test data set with $200\,{\upmu}{\rm m}$ Gaussian smearing.}
\end{figure}

In directional recoil detection, it is useful to break up directional performance into mean axial mismeasurement (the mean angle between the predicted axis and the true axis) and head/tail recognition efficiency (the fraction of tracks in which the head and tail ends of the recoil are predicted correctly)~\citep{Vahsen:2021gnb}. This is displayed for the $40\,$keV test data set with $443\,{\upmu}{\rm m}$ Gaussian smearing in Figure~\ref{svensplot1} and the $50\,$keV test data set with $200\,{\upmu}{\rm m}$ Gaussian smearing in Figure~\ref{svensplot2}. The vMF-NN and Det-NN models have similar performance at 100\% event efficiency, significantly outperforming the Non-ML models on both performance metrics. By omitting events with high predicted uncertainty from the vMF-NN model, performance can be further enhanced, at the cost of event efficiency.

Since our model is trained over a large range of diffusion and the simulated electrons include a significant amount of variability in terms of track shape and the amount of deposited ionization, we expect our model to be robust against modest variations in simulation parameters, including environmental conditions. We verify this with a test on varying temperature, which is the least tightly controlled environmental parameter in most setups. We create a new test set that is equivalent to the 40 keV test set with 443 um diffusion, but simulated at $22^\circ$C instead of $20^\circ$C. Even though the vMF-NN model is trained on $20^\circ$C simulations, the cosine distance loss it achieves on the $20^\circ$C and $22^\circ$C test sets agrees within 0.48\%. This indicates that small fluctuations in temperature have a very small systemic effect on our model.

\subsection{Comparison to Gauss-NN}
To highlight the advantages of the proposed vMF-NN model we compare to a second deep probabilistic model with the same architecture except it outputs the parameters of an isotropic 3D Gaussian distribution. This \textit{Gauss-NN} demonstrates the effect of using the wrong manifold for direction detection. Instead of a distribution over the sphere, it outputs a Gaussian p.d.f. over 3D space. The target direction (a point on the unit sphere) is modeled as a point in 3D space sampled from this distribution. The Gaussian is constrained to be isotropic in order to facilitate comparison to the vMF, so a single scalar parameter $\sigma$ controls the variance of the predicted distribution centered on $\hat{\mathbf{y}}_i$ (which is constrained to be on the sphere). Training minimizes the NLL, 
\begin{equation}
\label{NLL_Gauss}
    NLL_\textrm{Gauss} = \sum_i^N log(\sigma_i^3) + \frac{ (\mathbf{y}_i - \hat{\mathbf{y}}_i )^2 }{2 \sigma_i^2}.
\end{equation}

While this model is not a distribution over the sphere, it induces a distribution on the sphere known as the \textit{projected normal distribution} in directional statistics~\citep{wang2013directional}. Optimizing the NLL of the induced projected normal distribution is not practical because computing the p.d.f. requires integration along the radial direction, but the Gauss-NN objective can be viewed as \textit{approximating} that of the induced projected normal distribution on spherical data. Sampling from the induced projected normal distribution is trivial, and like the vMF-NN, cuts can be made on the $\sigma$ parameter to reject low-certainty samples. Thus, the Gauss-NN is an alternative to the vMF-NN with two key disadvantages: (1) computing the likelihood function over directions is computationally expensive, and (2) the approximate objective function contributes to miscalibrated uncertainty estimates.

To rigorously compare the performance of the deep learning models, we train each model five times with different random initializations and random mini-batches. The mean and standard deviation of the cosine distance loss on the two test sets are presented in Table~\ref{ensemble}. The two probabilistic models, vMF-NN and Gauss-NN, have nearly identical performance. The Det-NN has slightly lower error, but the difference is significant. The Best-Expected algorithm achieves the lowest error (but as we show above the vMF-NN can surpass it in performance with a 1\% efficiency cut).

\begin{table}[h!]
\centering
\begin{tabular}{|c| c|c |} 
 \hline
 Model Name & 40\,keV,   $443\,{\upmu}{\rm m}$  & 50\,keV,  $200\,{\upmu}{\rm m}$ \\ [0.5ex] 
 \hline
vMF-NN & $0.10562 \pm 0.00116$ & $0.06362 \pm 0.00064$ \\ 
Gauss-NN & $0.10540 \pm 0.00095$ & $0.06372 \pm 0.00088$ \\ 
Det-NN &  $0.10278 \pm 0.00104$ & $0.06236 \pm 0.00042$   \\ 
Best-Expected & $0.098 $ & $0.0569$  \\ 
 \hline
\end{tabular}
\caption{The mean and standard deviation of the cosine distance loss achieved by an ensemble of 5, for each of the models, on the $40\,$keV test data set with $443\,{\upmu}{\rm m}$ Gaussian smearing and the $50\,$keV test data set with $200\,{\upmu}{\rm m}$ Gaussian smearing. Since Best-Expected does not have any weights that need to be initialized, it performance is captured by a single value.}
\label{ensemble}
\end{table}

\begin{figure}[ht]
\begin{center}
\includegraphics[width=0.70\textwidth,trim={0.55cm 0.55cm 1cm 0.25cm},clip]{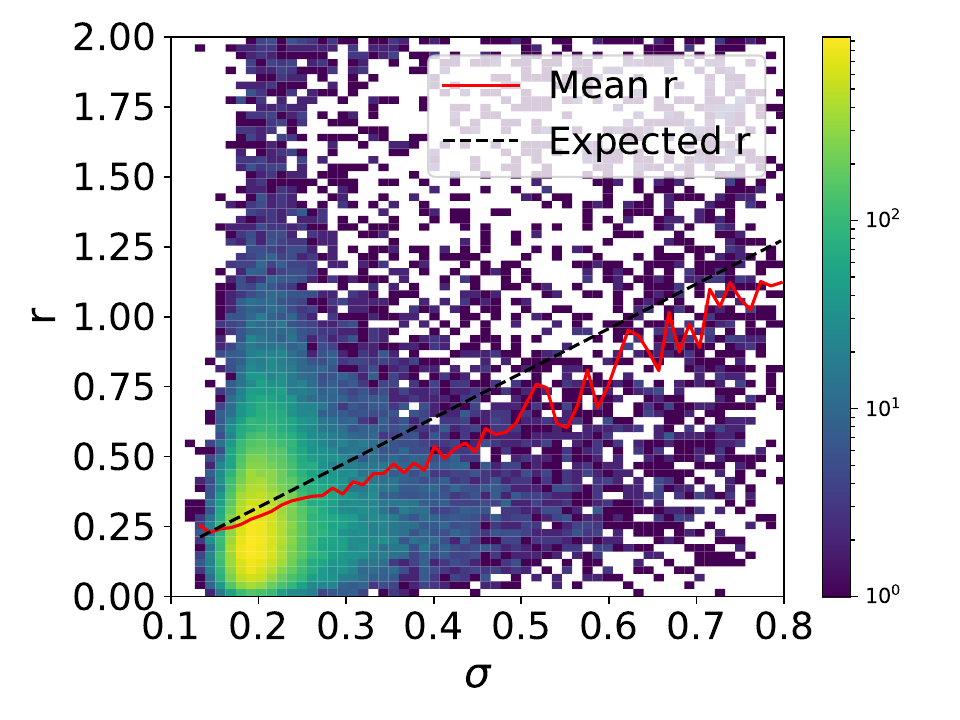}
\end{center}
\caption{
2D histogram of predicted uncertainty in the Gauss-NN ($\sigma$) and the Euclidean distance ($r$) between predicted and target directions for all test data. The red curve is the mean $r$ for each bin in $\sigma$. The black dashed curve is the expectation when $r$ is sampled from a 3D Gaussian with standard deviation $\sigma$. The systematic bias demonstrates a miscalibration of the Gauss-NN model.}
\label{Gauss-calib}
\end{figure}

\begin{figure}[ht]
\begin{center}
\includegraphics[width=0.70\textwidth,trim={0.35cm 0.35cm 0.35cm 0.35cm},clip]{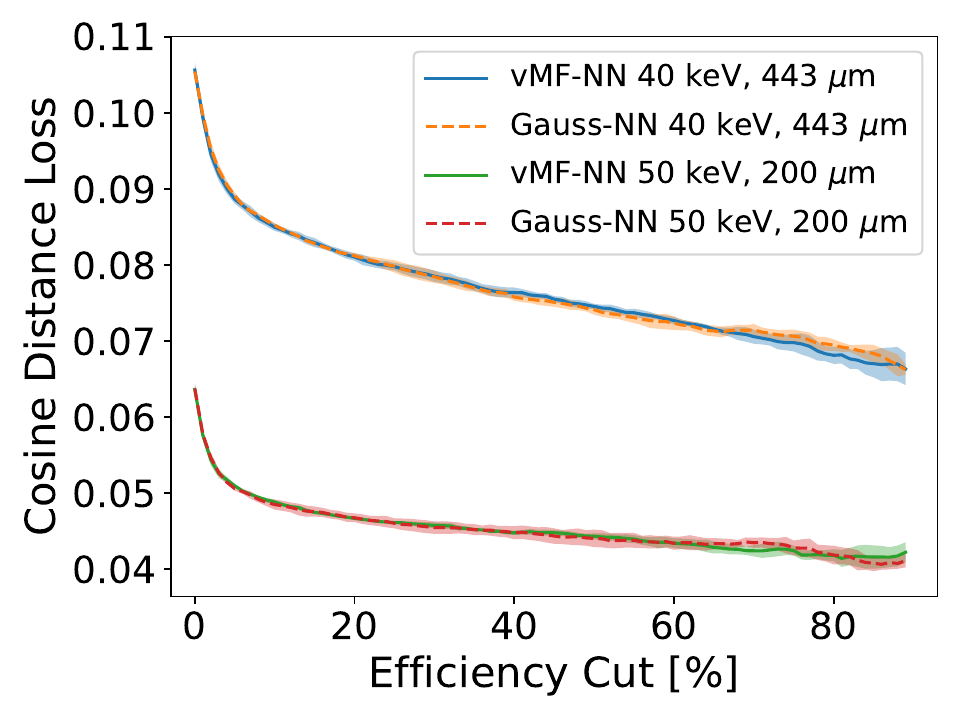}
\end{center}
\caption{
Cosine distance loss for the Gauss-NN and vMF-NN models at different efficiency cuts. Five models of each type were trained on the $40\,$keV test data set with $443\,{\upmu}{\rm m}$ Gaussian smearing and the $50\,$keV test data set with $200\,{\upmu}{\rm m}$ Gaussian smearing. The mean and standard deviation of the test losses are plotted versus efficiency cuts where the $\sigma$ and $\kappa$ parameters are used to remove samples with high predicted uncertainty for the Gauss-NN and vMF-NN models, respectively.
}
\label{Gauss-compare}
\end{figure}

The fact that the homoscedastic model performs slightly better than the heteroscedastic models is not surprising. This phenomena has been studied in the machine learning literature~\citep{NEURIPS2019_07211688,seitzer2022on} and there exist methods to reduce this gap~\citep{pmlr-v206-stirn23a,NEURIPS2023_a901d554}. The two probabilistic models performing similarly is explained by the models having identical architectures and very similar objective functions (Equation~\ref{NLL} and Equation~\ref{NLL_Gauss}). The main differentiator between the vMF-NN and Gauss-NN is the calibration --- while the vMF-NN is shown to be well calibrated in Figure~\ref{2dhist}, the Gauss-NN suffers from systematic miscalibration in which the variance is overestimated (Figure~\ref{Gauss-calib}). This is explained by the mismatch between the topology of the probabilistic model (in 3D) and the data (on the sphere). Together with the requirement of computing an integral to obtain the likelihood function for the induced projected normal distribution, this makes the Gauss-NN model a poor alternative compared to the vMF-NN for applications in which accurate calibration is needed. However, Figure~\ref{Gauss-compare} shows that the Gauss-NN performs just as well on our application, where only the relative ordering of the uncertainties is important for effectively rejecting low-certainty samples.

\section{Discussion}

This work integrates all the essential components required for applying deep learning to modern directional recoil detectors. We describe a deep probabilistic approach for predicting 3D direction with well-calibrated uncertainties that can be used to reject low-certainty events. Using the design patterns of deep learning, we developed a 3D convolution approach that processes high-dimensional event data from Gas TPC detectors. Together, these contributions provide a probabilistic deep learning approach that advances the state-of-the-art for analyzing  directional recoil detector data. Moreover, the method is applicable to any task requiring 3D direction estimation with uncertainty.

The benefits of obtaining directional information are profound and could prove game-changing in several areas of fundamental physics. Most importantly, dark matter detection experiments could reject solar neutrino backgrounds based on recoils pointing back towards the sun and positively identify dark matter from the observed angular distribution of recoils~\citep{Billard:2013qya, OHare:2021utq}. This would overcome two major limitations in current experiments seeking to identify the particle nature of dark matter, which is arguably the most urgent problem in contemporary physics. Directional measurements of electron recoils induced by solar neutrinos would enable improved measurements of neutrinos from the Sun's ``CNO cycle'' which could settle a long-standing puzzle known as the solar abundance problem~\citep{ villante2019updated,OHare:2022jnx,Lisotti:2024fco}. Directional measurements of photoelectrons produced by X-rays enable us to map the X-ray polarization of extended astrophysical sources, as in the Imaging X-ray Polarimetry Explorer ~\citep{WEISSKOPF20161179}. A comprehensive review of directional recoil detection and its benefits can be found in ~\cite{Vahsen:2021gnb}.

\section{Conclusion}

For the first time, we describe a deep probabilistic approach for predicting 3D direction. The model utilizes the von Mises-Fisher to predict distributions on $\mathbb{S}^2$ and we specify approximations of the NLL that enable stable training. Two tests are developed for assessing model calibration in direction prediction applications and the model is shown to be well-calibrated in our experiments.

The approach is demonstrated on an application to directional recoil detectors in physics. The model is compared to a variety of alternative approaches: a non-ML approach, a deterministic ML model without uncertainty, and a probabilistic deep learning model that does not account for the topology of the data. The results show that the ML models drastically out-perform the non-ML approach. While all the ML approaches perform very similarly in terms of cosine distance error, the probabilistic ML approach is shown to realize significantly lower errors by rejecting a small set of low-certainty events. Finally, it is argued that deep probabilistic direction prediction models that do not account for the true topology of the data have practical disadvantages and are less well-calibrated.

\subsection*{Acknowledgments}
MG thanks Samuele Torelli for useful discussions on direction fitting for X-ray polarimetry. We thank Jeffrey Schueler for providing the code used to render the electron recoil in Figure 1. MG and SEV acknowledge support from the U.S. Department of Energy (DOE) via Award Number DE-SC0010504. PS acknowledges support from the National Science Foundation under Grant 2238375. The technical support and advanced computing resources from University of Hawaii Information Technology Services – Cyberinfrastructure, funded in part by the National Science Foundation CC* awards 2201428 and 2232862 are gratefully acknowledged.

\bibliographystyle{abbrvnat}
\bibliography{mybib}

\appendix

\section{Non-Machine Learning algorithms}
\label{NML_algo}

Here, we detail the non-machine learning algorithms to which we compare our models. The first algorithm, Non-ML, is adapted from~\cite{DiMarco_2022} where it is discussed in the context of determining the initial direction of a photoelectron track imaged in 2D by the Imaging X-ray Polarimetry Explorer (IXPE)~\citep{WEISSKOPF20161179}. We generalized the algorithm to 3D so that it can be applied to our simulations. The Non-ML algorithm is outlined below:
\begin{enumerate}
    \item Find the barycenter ($\mathbf{x}_b$) of the track, defined as $\mathbf{x}_b = \frac{\sum_i q_i  \mathbf{x}_i }{ \sum_i q_i}$ , where $q_i$ is the charge in a voxel and $\mathbf{x}_i$ is the position of the voxel. Center the track on $\mathbf{x}_b$.
    \item Compute the weighted covariance matrix. Use its singular value decomposition (SVD) to find the principal axis $\mathbf{v}_{\textrm{PA}}$ (the axis on which the second moment is maximized) and the second moment along $\mathbf{v}_{\textrm{PA}}$, denoted as $M_2$.
    \item Compute the third moment ($M_3$) about $\mathbf{v}_{\textrm{PA}}$. $M_3$ is also known as the skewness and it is used to keep only the initial part of the recoil track. Keep only the points satisfying $sgn( (\mathbf{x}_i-\mathbf{x}_b) \cdot \mathbf{v}_{\textrm{PA}}) = sgn(M_3)$.
    \item $M_2$ sets the length scale along the principal axis, keep only the points satisfying $ 1.5 M_2 < (\mathbf{x}_i-\mathbf{x}_b) \cdot \mathbf{v}_{\textrm{PA}} < 3 M_2$
    \item The two conditions above are meant to isolate the beginning portion of the track. The interaction point ($\mathbf{x}_{\textrm{IP}}$) is defined as the charge-weighted center of the remaining points. Center the remaining points on $\mathbf{x}_{\textrm{IP}}$.
    \item The remaining points are re-weighted with $w_i = \exp{\left( -dist(\mathbf{x}_{\textrm{IP}},\mathbf{x}_i)/w_o \right)}$, where $w_o = 0.05$ as specified in~\cite{DiMarco_2022}. Compute the weighted covariance matrix and use SVD to find the principal axis, this gives the initial axial direction of the track, denoted as $\mathbf{v}_\textrm{IP}$.
\end{enumerate}
Above, $1.5 M_2$, $3 M_2$, and $w_o = 0.05$ are all adopted from~\cite{DiMarco_2022}, it is possible that these values are not optimal for our application. Hence, we also investigate the performance when these parameters are tuned to minimize the cosine distance loss on each test data set. There is no discussion of assigning a head-tail to $\mathbf{v}_\textrm{IP}$. We employ two approaches: In the first, we assign the head-tail such that $ \mathbf{v}_\textrm{IP} \cdot \mathbf{v}_\textrm{PA} \geq 0$. In the Second, we simply assign the correct head-tail, such that $ \mathbf{v}_\textrm{IP} \cdot \mathbf{v}_\textrm{True} \geq 0$ where $\mathbf{v}_\textrm{True} $ is the true direction. Finally, the selections applied in steps $3$ and $4$ may not leave enough points to determine a principal axis for a subset of the tracks. In this case, the track is omitted and the track efficiency is noted. When tuning the parameters, we only consider cases resulting in a track efficiency of greater than $10\%$.

The second algorithm, Best-Expected, attempts to estimate the best possible directional performance by using information that is only available in simulation. The purpose is not to be a viable option for analysis but to provide a benchmark allowing us to check how close our models are to the best expected performance. The implementation of this method is outlined below:
\begin{enumerate}
    \item For each simulation, Identify the true starting point. The true starting point is information that is only known in simulation. One of the key challenges faced by electron direction finding models is identifying the starting point.
    \item Make a sphere of radius $\varepsilon$ centered on the true starting point. The value of $\varepsilon$ is specified by tuning it to minimize the cosine distance loss for each test data set.
    \item Determine the principal axis of the points contained within the sphere using SVD.
    \item Assign a head-tail to the principal axis such that it agrees with the true direction. In this step, the true label is used to assist the algorithm, giving it a significant advantage. 
    \item The final vector direction obtained is used as the predicted direction of this method. 
\end{enumerate}
Since this algorithm is given information about the true starting point, it is tuned on the test data sets, and it uses the truth head-tail information, we reason it is a good measure of the best expected performance.

\section{Application to a Toy Problem}
\label{arrows}

\begin{figure}[ht]
\begin{center}
\includegraphics[width=0.75\textwidth,trim={1cm 1cm 0cm 2cm},clip]{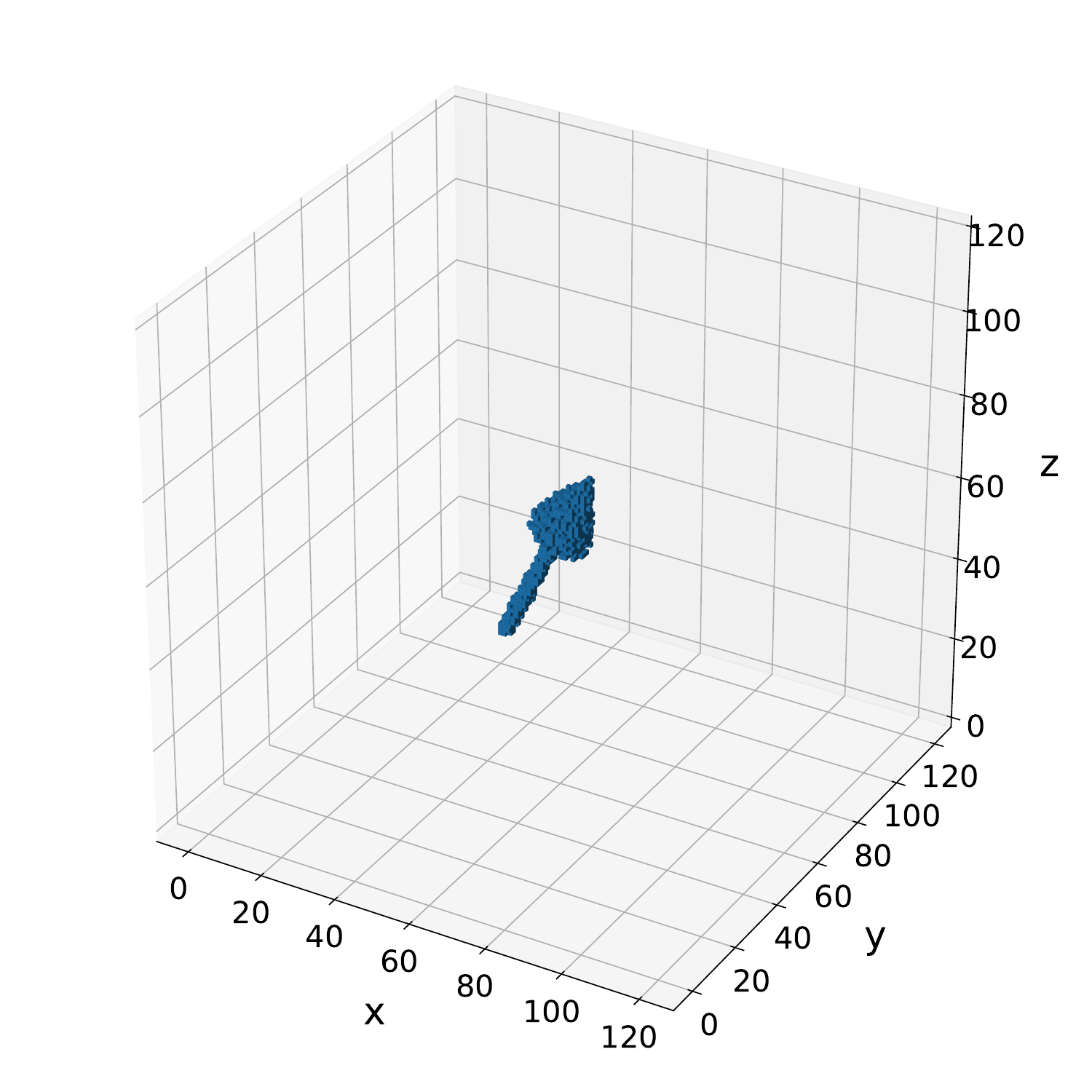}
\end{center}
\caption{An example of a 3D arrow for the application of our models to a simple toy case.}
\label{arrow_example}
\end{figure}

In this section, we provide an illustrative example of our method on a toy problem of determining the direction of 3D arrows. Arrows with a random direction, fixed shape, and uniform density are generated and binned into a $(120,120,120)$ voxel grid. An example is displayed in figure~\ref{arrow_example}. We generate $6000$ arrows, $4000$ are used for the training set and $1000$ are used for the validation and test sets. The Det-NN and vMF-NN models are trained on the training set while their validation loss is used for early stopping. If the validation loss has not decreased in the last 2 epochs, training is stopped. The weights corresponding to the lowest validation loss are saved and used for the final model.

In this simple case, both models are fully trained within minutes. The cosine distance loss of the Det-NN and vMF-NN models on the test data set is $4 \times 10^{-4}$ and $8 \times 10^{-4}$, respectively. As expected, the direction predictions are much more accurate than Section~\ref{e_recoils}. The lowest predicted $\kappa_i$ by vMF-NN on the test data set is $443$, meaning the model is much more confident in the predicted directions than for the electron recoil case. The code needed to create the arrow data sets, train our models, and test our models is available at~\url{https://github.com/mghrear/3D_Heteroscedastic_Convnet}.

\end{document}